**Diagnostic Impact of Cine Clips for Thyroid Nodule Assessment on Ultrasound**

Jichen Yang[a], Brian C. Allen[b], Kirti Magudia[b], Lisa M. Ho[b], Chad M. Miller[b], Maciej A. Mazurowski[a,b], Benjamin Wildman-Tobriner[b]

[a] Department of Electrical & Computer Engineering, Duke University, 130 Hudson Hall, Box 90291, Durham, NC 27708, USA

[b] Department of Radiology, Duke University Medical Center, 2301 Erwin Road, Box 3808, Durham, NC 27710, USA

Corresponding Author:

Jichen Yang,

jy168@duke.edu,

Duke University, Department of Electrical and Computer Engineering

Duke University Pratt School of Engineering

Box 90291

Durham, NC 27708, USA

+1 984-244-9678

Running title: Impact of Cine in Thyroid Nodule Diagnosis

**Abstract**

**Objective:** Thyroid ultrasound is commonly performed using a combination of static images and cine clips (video recordings). However, the exact utility and impact of cine images remains unknown. This study aimed to evaluate the impact of cine imaging on accuracy and consistency of thyroid nodule assessment, using the American College of Radiology Thyroid Imaging, Reporting and Data System (ACR TI-RADS).

**Methods:** 50 benign and 50 malignant thyroid nodules with cytopathology results, from 89 patients, were included. A reader study with 4 specialty-trained radiologists was then conducted over 3 rounds, assessing only static images in the first two rounds and both static and cine images in the third round. TI-RADS scores and the consequent management recommendations were then evaluated by comparing them to the malignancy status of the nodules.

**Results:** Mean sensitivity for malignancy detection was 0.65 for static images and 0.67 with both static and cine images ($p=0.41$). Specificity was 0.20 for static images and 0.22 with both static and cine images ($p=0.25$). Management recommendations were similar with and without cine images. Intrareader agreement on feature assignments was slightly lower when cine images were added (0.83 vs 0.87, $p<0.001$), and TI-RADS point totals changed more between rounds with cine images.

**Conclusion:** The inclusion of cine imaging for thyroid nodule assessment on ultrasound did not significantly change diagnostic performance in this reader study, although it changed some feature assignments. Current practice guidelines, which do not mandate cine imaging, may be sufficient for TI-RADS–based management recommendations.

## Introduction

Due to its cost-effectiveness, wide availability, and high anatomic detail for superficial structures, ultrasound (US) remains the first-line imaging modality for evaluation of thyroid nodules.[1] During assessment of a nodule on US, sonographers may obtain and radiologists may subsequently interpret a combination of static grayscale images in addition to color Doppler images.[2] Another tool that may be employed during nodule evaluation is cine imaging.[3] To acquire a cine (also called a cine clip or video clip), the sonographer sweeps the field of view across an area of interest while recording, giving radiologists a moving picture of an abnormality and its immediate surroundings. As a result, cine clips may offer a more comprehensive assessment of what a sonographer viewed in real-time, as opposed to static images that provide limited data.

Despite potentially adding imaging information, the precise value of cine clips for thyroid nodules on US remains unknown. The American College of Radiology Thyroid Imaging, Reporting and Data System (ACR TI-RADS), a commonly implemented system for nodule classification and risk stratification, states in its white paper that "ultrasound video clips are valuable to provide further information about the spatial relationships between nodules and adjacent structures."[4] Indeed, some experts argue that cines are critical, allowing for better assessment of nodule features, for example showing the composition of a nodule in its entirety instead of seeing only a portion on a single image.[5,6] Nonetheless, ACR TI-RADS does not mandate cine clips, and some practices do not use them for thyroid US, likely due to time pressures on both sonographers and radiologists. Beyond risk stratification, cine clips can help identify a pyramidal lobe,[7] are used in the follow-up of differentiated thyroid cancer,[8] and, as tracked sweeps, allow three-dimensional volume measurement.[9] Only a few studies have directly compared cine clips with static images for thyroid nodules, and they were small or limited to a single reader.[10–12]

Consequently, the purpose of our study was to examine the impact of cine imaging on the evaluation of thyroid nodules on US. We hypothesized that the addition of cine images to US examinations would allow for more accurate and consistent diagnosis. Specifically, we aimed to look at how cine images impacted ACR TI-RADS scores, risk categories, and diagnosis of malignancy (via recommendation for fine needle aspiration). These data could help improve practice guidelines moving forward.

## Materials and Methods

This retrospective study was institutional review board (IRB)-approved (Duke University IRB #00081895 including waiver of informed consent) and Health Insurance Portability and Accountability Act compliant. It was conducted in accordance with the Declaration of Helsinki.

### *Study Population*

Our institution's pathology report database was searched from 2017-2020 to identify a cohort of thyroid nodules that had cytopathologic results based on fine needle aspiration (FNA). Rather than include consecutive studies, we designed our cohort with 50% benign nodules (Bethesda II) and 50% malignant or suspicious for malignancy nodules (Bethesda V and VI).[13] We took this approach in order to include a variety of US appearances and features. Because of the low prevalence of malignant nodules, using consecutive nodules would have resulted in a predominantly benign population and possibly a limited number of US features. Since the study goal was to assess the impact of cine imaging on reader behavior and performance, the prevalence of disease (and its potential effects on positive and negative predictive value) was not a focus.

Over this time period at our institution, cine clips were not routinely performed on every thyroid US, so nodules with Bethesda V/VI cytopathology were manually reviewed by a subspecialty trained radiologist to identify those with cine clips of the nodule. The radiologist also selected relevant transverse and longitudinal static images of the nodule. This radiologist did not participate as a reader. Ultimately 50 malignant cases were identified. Next, over the same time period, a random sampling of 50 benign nodules with cines were also selected. Their static images were also selected independently of the readers. Nodules were chosen randomly to ensure the benign nodules were distributed across the same time period rather than being clustered, to avoid any bias based on equipment or staffing. 50 benign nodules with cine images were chosen, and the final cohort consisted of 100 nodules, each with static images and cine clips of the nodule in question. Ultrasound examinations were performed with a variety of commercially available units (iU22 and ATL HDI 5000, Philips, Amsterdam, Netherlands; LOGIQ E9, General Electric, Boston, MA, USA) equipped with 5–15 MHz linear array transducers.

*Reader Study*

Four subspecialty trained radiologists with routine practice in interpreting thyroid ultrasounds participated in the study (experience ranging from 1-25 years). Each reader independently completed three rounds of US reads. For the first round, readers were given only static US images of each nodule (one transverse and one longitudinal view, focused on the center of the nodule as recorded by the sonographer) and asked to assign TI-RADS features to every nodule. After a washout period of approximately 12 weeks, readers completed a second round that was the same as the first, only with a new randomized order. This repeat round was included to assess intrareader variability. After another washout period of approximately 12 weeks, the third and final round provided readers with static images and cine clips for assessment, which simulated the current standard of care. Two cine clips of each nodule were provided in this round, and readers were not required to view them.

TI-RADS scores for all readers and all rounds were automatically tabulated and correlated with nodule size according to the TI-RADS algorithm. As specified by ACR TI-RADS, nodules rated spongiform received 0 points in total,[4] and a point total of 1, which the ACR TI-RADS chart does not list, was assigned to TI-RADS level 2 (TR2). Recommendations for no follow-up, follow-up, or FNA were calculated. Performance for detection of malignancy based on FNA recommendations was then calculated using sensitivity and specificity. Intrareader variability between rounds 1 and 2 was calculated to help determine if any changes detected with and without cine clips were due to the cine clips or inherent reader variability due to a repeated read at a later time. Finally, performance with and without cine clips was compared.

*Machine Learning*

To better evaluate the performance of four readers and to estimate the overall difficulty of the dataset, we deployed a previously developed machine learning algorithm that was designed to differentiate benign from malignant thyroid nodules. This algorithm used a convolutional neural network, which was trained on 1278 thyroid nodules from an earlier cohort at our institution (2006–2010). It achieved 0.87 sensitivity and 0.52 specificity during previous testing.[14] For the current study, the algorithm was run on the static images of the 100 nodules. Its malignancy predictions were converted to risk levels using the previously published thresholds and combined with ACR TI-RADS size criteria to generate FNA recommendations, and its area under the receiver operating characteristic curve (AUC) was also calculated.

*Statistical Analysis*

The performance for each reader in each round was evaluated by calculating sensitivity and specificity. Differences in management decision were evaluated by calculating the number of cases where a reader gave different FNA recommendations to the same case between rounds. TI-RADS point differences were evaluated by calculating the average of the absolute total point differences between each round for each case for all readers. Agreement of each TI-RADS feature assignment was evaluated by calculating the proportion of nodules that were assigned the same feature category by the same reader in each pair of rounds. Intrareader agreement was also summarized with Cohen's κ, and interreader agreement among the four readers in each round with Fleiss' κ.[15,16] Discrimination by TI-RADS point totals was summarized with the AUC and compared between rounds with the Obuchowski–Rockette method for multireader multicase data, with Hillis degrees of freedom and both readers and nodules treated as random effects.[17,18] Sensitivity and specificity of TI-RADS levels without size criteria (TR4 or higher, and TR5) were also calculated, and the sizes of benign and malignant nodules were compared with the Mann–Whitney U test. Descriptive 95% confidence intervals (CIs) for individual readers and for reader averages were obtained by bootstrap resampling of nodules. Because some patients contributed more than one nodule, 95% CIs and two-sided p-values for comparisons between rounds were obtained by bootstrap resampling of patients, with the four readers held fixed (10,000 resamples in both bootstraps).[19] These comparisons were between the still-only rounds (rounds 1 and 2, averaged) and the cine-included round (round 3) or, for changes and agreement between rounds, between the still-only pair of rounds and the two pairs that included round 3 (averaged). The primary comparison was the difference in mean sensitivity and specificity between still-only and cine-included rounds. Per-reader changes in sensitivity and specificity between rounds 2 and 3 were assessed with exact McNemar tests. No power calculation was performed; we report 95% CIs for the differences instead.[20] Comparisons of the frequencies of the 21 individual feature categories were exploratory and were not adjusted for multiple comparisons. Lastly, sensitivity and specificity were compared to results from the deep learning algorithm.

P-values of less than 0.05 were considered significant, and computations were performed in Python with the NumPy, SciPy, pandas, statsmodels, and scikit-learn packages; our Obuchowski–Rockette implementation reproduced the results of the OR-DBM MRMC software for a published example dataset. A large language model was used to help write and run the analysis code; the authors reviewed the analyses and take full responsibility for them.

**Results**

Average nodule size was 2.4 ± 1.4 cm. The 100 nodules were from 89 patients, 10 of whom contributed more than one nodule. Benign nodules were larger than malignant nodules (2.7 ± 1.2 vs 2.2 ± 1.6 cm, p=0.004), and 21 of 50 malignant nodules (42%) but only 5 of 50 benign nodules (10%) measured less than 1.5 cm. Nodule features, as reported by readers across all three rounds, are shown in Table 1. All TI-RADS features were represented in the dataset.

Several feature assignments differed in the cine-included round. Macrocalcifications were reported in 23.5% of assessments with cine images, compared with 17.0% and 16.3% in the two still-only rounds (p<0.001), and peripheral (rim) calcifications in 10.5%, compared with 7.0% and 8.0% ($p=0.02$), whereas very hypoechoic echogenicity (10.0%, compared with 13.8% and 14.0%; $p=0.001$) and ill-defined margins (17.5%, compared with 23.0% and 29.3%; $p<0.001$) were assigned less often. Margin assignments also differed between the two still-only rounds (Table 1).

Across the four readers, mean sensitivity for detection of malignancy was similar between still-only rounds and cine-included (0.645 vs 0.670; difference, 0.025 [95% CI: −0.031, 0.078]; $p=0.41$). Specificity was also similar, with a mean of 0.195 for still-only and 0.220 after the addition of cine images (difference, 0.025 [95% CI: −0.015, 0.068]; $p=0.25$). No individual reader's sensitivity or specificity changed significantly between rounds 2 and 3 (exact McNemar $p>0.2$). As a comparator for overall dataset difficulty, the machine learning algorithm achieved a sensitivity of 0.90 and a specificity of 0.04 (based on still-only images), because it assigned 87 of 100 nodules to its highest risk level; its AUC was 0.51. Complete reader sensitivities and specificities are found in Table 2.

The number of cases that had different management decisions between still-only rounds ranged from 9 to 13, with an average of 11. When comparing still-only to cine rounds, those differences were similar, with readers changing recommendations between 8 and 15 cases, with an average of 12.5 ($p=0.38$). More detailed information on changes in management recommendations between rounds is outlined in Table 3.

When looking at point differences between rounds, the average absolute TI-RADS points difference between still-only rounds for all readers ranged from 0.98 to 1.21, with an average of 1.07. The average absolute TI-RADS points difference between still-only rounds and cine-included rounds for all readers ranged from 1.02 to 1.93, with an average of 1.39 (p-value $<$ 0.001 comparing to still-only rounds). Additional details of point differences are shown in Table

4. These changes occurred in both directions, and mean point totals were similar in still-only and cine-included rounds (5.60 vs 5.71, p=0.35). Between still-only and cine-included rounds, point totals changed in 479 of 800 paired assessments; 299 (62%) of these changes moved the nodule to a different TI-RADS level, and only 100 (33%) of the level changes altered the FNA recommendation, because nodule size often met, or failed to meet, the FNA threshold at both levels. Between the two still-only rounds, point totals changed less often (205 of 400 paired assessments), but these two proportions were similar (63% and 34%).

At the feature level, readers showed consistency between still-only rounds compared to the cine-included round. For example, the agreement of composition between still-only rounds for all readers ranged from 0.87 to 0.95, with an average of 0.92. The agreement of composition between still-images-only round and cine-included round for all readers ranged from 0.86 to 0.93, with an average of 0.90. Agreement for the remaining features including echogenicity, shape, margin, and echogenic foci followed similar trends, with complete data shown in Table 5. Averaged across all features and readers, agreement was 0.87 between the still-only rounds and 0.83 between still-only and cine-included rounds (difference, −0.039 [95% CI: −0.053, −0.026]; p<0.001), with the largest decreases for macrocalcifications and punctate echogenic foci. For every feature, mean intrareader Cohen's κ was lower between still-only and cine-included rounds (range, 0.39 to 0.71) than between the two still-only rounds (Table 6). Interreader agreement (Fleiss' κ) was higher for the FNA recommendation (0.64, 0.63, and 0.56 in rounds 1–3) than for the TI-RADS level (0.32, 0.30, and 0.29).

The reader-averaged AUC of TI-RADS point totals was 0.623, 0.628, and 0.645 in rounds 1, 2, and 3, respectively, and did not differ significantly between still-only and cine-included rounds (Table 7). Table 7 also shows the sensitivity and specificity of TI-RADS levels used without size criteria.

## Discussion

Our study of four radiologists interpreting thyroid US both without and with the use of cine imaging showed overall very little change with the addition of cine images. Overall performance was not significantly affected, management recommendations were similar, and feature assignments were largely consistent. While points were assigned slightly differently between rounds, these changes did not significantly affect ultimate diagnostic performance. Cine images did, however, change how some features were assigned, particularly macrocalcifications.

To our knowledge, the impact of cine images on thyroid US performance has not been comprehensively studied. Intuitively, certain TI-RADS features seem like they might be better assessed by cine images, for example echogenic foci that may only be present in part of a nodule or margins that might be hard to appreciate on a static, two-dimensional image. However, our results suggest that using cines may not capture these theoretical benefits, with no significant downstream change in diagnostic performance. The underlying reasons for these results are likely complex and multifactorial, but a few factors may be considered. First, it is possible that readers could assign features well enough using static images alone, and that cine images simply didn’t add value. The representative static images are generally the largest diameter and central portion of the nodule and typically show margins fairly well. Scenarios in which punctate echogenic foci or macrocalcifications were incompletely evaluated on static images may have been rare, and other features may have been discernable enough on still images. Our feature-level results only partly support this explanation, because readers reported macrocalcifications more often and ill-defined margins less often with cine images. Second, it is possible that despite having cines available in the final round, readers still made decisions using still images since both were provided. We provided cines for the final round but did not mandate that readers view them. Some readers may simply prefer static images, even if cines are available. Nonetheless, the shifts in several feature assignments, the larger point changes, and the lower agreement with the still-only reads in the cine-included round suggest that readers did use the cines. Similar findings have been reported by others, with macrocalcifications identified more often on cine clips[12] and lower interobserver agreement on cine clips than on static images.[10] Third, the static images of the malignant nodules in our study were selected by a subspecialty trained radiologist and may have shown key features more completely than images chosen by less experienced users;[10] cine clips may add more value in that setting.

These results may be informative on several levels. First, there exist potential practice pattern ramifications, as some radiologists (as well as endocrinologists and surgeons) routinely use cine images, while others do not. While we do not suggest that cines be eliminated from practice, our results suggest that those who do not use them (whether for time savings, user preference, or otherwise) are still likely meeting the standard of care for TI-RADS risk stratification. A second implication relates to the large body of artificial intelligence (AI) work that is currently developing for thyroid nodules. Hundreds of papers and a handful of commercially-available products have been produced for AI analysis of thyroid nodules, and the large majority of them are founded on (and function using) still images.[14,21,22] Some commercially-available products do accept cine clips,[23,24] and research models that analyze cine clips or videos have been developed,[25] some reporting better performance than models based on single images.[26,27] Using cine images for more complete nodule characterization stands as a logical next step in these efforts, but our results question whether that research and development time will result in any meaningful advances. Our study assessed radiologists rather than algorithms, which may use information in cine clips that human readers do not.

This study had several limitations, some of which warrant discussion. In particular, overall reader performance was well below expected values for radiologists interpreting thyroid US with TI-RADS. To troubleshoot this result, we used our group's deep learning algorithm that has been shown to characterize thyroid nodules as well as radiologists.[14] The algorithm's performance was also below its expected result (AUC, 0.51), suggesting that this dataset was a challenging one, although differences between the algorithm's training images and our images may also have contributed. We are uncertain as to why reader performance was poor, though it may be a result of selection bias related to finding scans that had cines. Our study period took place before routine cines for every nodule, so studies with cines may include inherently larger or more challenging nodules that prompted a sonographer to acquire a cine. Nodule size and case mix also likely contributed. All nodules had undergone FNA, and the benign nodules in our cohort were larger than the malignant nodules, so ACR TI-RADS size thresholds led to FNA recommendations for many benign nodules and withheld FNA for many small malignant nodules. The enriched prevalence of malignancy may also have lowered specificity.[28] Reader sensitivity was similar to that reported in a large multi-institutional cohort (68%),[29] but specificity was lower than reported in other studies of biopsied nodules (27%–53%).[26,29,30] Specificity remained low when TI-RADS levels were used without size criteria (30% in the still-only rounds and 35% with cine images for TR4 or higher [Table 7], compared with a pooled estimate of 55%).[31] Nonetheless, the results comparing still-only reads to cine-included reads still stand, and we feel the low performance does not render the results invalid. Beyond low performance, another limitation is that we did not measure reader behaviors that may also be of interest. For example, we did not time the interpretations, a metric which may show differences with the addition of cines. Nor did we ask about reader confidence or preference, metrics that might not directly impact performance, but are nonetheless important. Because viewing of the cine clips was not required and we did not include a cine-only round, we cannot determine how much each reader relied on the cines. The cine-included round was always performed last, so effects of reading order or time cannot be fully excluded. Our reference standard was FNA cytopathology rather than surgical histopathology, and nodules with indeterminate cytopathology (Bethesda III and IV) were not included; because Bethesda categories do not perfectly predict final histopathology,[13] some nodules may have been misclassified, which would tend to make differences between rounds smaller. With 100 nodules, the study could not exclude improvements of up to 0.08 in sensitivity or 0.07 in specificity with cine images (upper 95% confidence limits). In addition, we did not assess other uses of cine clips, such as detection of a pyramidal lobe or volumetric assessment. Patient sex was not collected, so results could not be examined by sex. Finally, our study used radiologists from a single center, all of whom likely interpret thyroid US similarly. Asking radiologists from another practice may yield different results.

## Conclusion

In conclusion, the addition of cine images to interpretation of thyroid nodules on US did not have a significant impact on reader performance for TI-RADS risk stratification in this study, although it changed some feature assignments. These data may be helpful for practices deciding on their acquisition and interpretation protocols.

**Authorship Contribution**

J.Y. prepared the data, executed the reader study, performed statistical analysis, deployed the machine learning model, drafted the manuscript

B.C.A., K.M., L.M.H. & C.M.M. participated in the reader study, edited the manuscript

M.A.M. & B.W.T. conceived the study, provided guidance on experimental design, supervised the research, and edited the manuscript

**Acknowledgments**

This research did not receive any specific grant from funding agencies in the public, commercial, or not-for-profit sectors.

**Conflict of Interest Statement**

J.Y., B.C.A., K.M., L.M.H., C.M.M. & M.A.M. declared no conflict of interest.

B.W.T. is a consultant for See-Mode Technologies.

**Data Availability Statement**

The data supporting the findings of this study are not publicly available because they contain patient information; de-identified data are available from the corresponding author on reasonable request.

**Declaration of Generative AI and AI-assisted technologies in the writing process**

During the preparation of this work the authors used a large language model in order to help write and run the analysis code and to help refine and condense the manuscript text. After using this tool/service, the authors reviewed and edited the content as needed and take full responsibility for the content of the publication.

## References

1. Cao C-L, Li Q-L, Tong J, et al. Artificial intelligence in thyroid ultrasound. Front Oncol 2023;13:1060702; doi: 10.3389/fonc.2023.1060702.

2. Chaudhary V, Bano S. Thyroid ultrasound. Indian J Endocrinol Metab 2013;17(2):219–227; doi: 10.4103/2230-8210.109667.

3. Seifert P, Maikowski I, Winkens T, et al. Ultrasound cine loop standard operating procedure for benign thyroid diseases—evaluation of non-physician application. Diagnostics (Basel) 2021;11(1):67; doi: 10.3390/diagnostics11010067.

4. Tessler FN, Middleton WD, Grant EG, et al. ACR Thyroid Imaging, Reporting and Data System (TI-RADS): white paper of the ACR TI-RADS Committee. J Am Coll Radiol 2017;14(5):587–595; doi: 10.1016/j.jacr.2017.01.046.

5. Low G, Bara M, Du Y, et al. Tips for improving consistency of thyroid nodule interpretation with ACR TI-RADS. J Ultrason 2022;22(88):e51–e56; doi: 10.15557/JoU.2022.0009.

6. Du Y, Bara M, Katlariwala P, et al. Effect of training on resident inter-reader agreement with American College of Radiology Thyroid Imaging Reporting and Data System. World J Radiol 2022;14(1):19–29; doi: 10.4329/wjr.v14.i1.19.

7. Yan Y, He X, Hu K, et al. Cine clips increase the detection of thyroid pyramidal lobe in routine thyroid sonogram. Ultrasound 2024;32(3):157–163; doi: 10.1177/1742271X231225047.

8. Sopuschek MP, Freesmeyer M, Winkens T, et al. Standard operating procedure (SOP) for cervical ultrasound cine loop video sequences in the follow-up of differentiated thyroid carcinoma (DTC). Endocrine 2025;87(2):635–647; doi: 10.1007/s12020-024-04021-w.

9. Krönke M, Eilers C, Dimova D, et al. Tracked 3D ultrasound and deep neural network-based thyroid segmentation reduce interobserver variability in thyroid volumetry. PLoS One 2022;17(7):e0268550; doi: 10.1371/journal.pone.0268550.

10. Schenke SA, Petersen M, Görges R, et al. Interobserver agreement in ultrasound risk stratification systems for thyroid nodules on static images versus cine-loop video sequences. Diagnostics (Basel) 2024;14(19):2138; doi: 10.3390/diagnostics14192138.

11. Schmidt TN, Freesmeyer M, Kühnel C, et al. Risk stratification of thyroid nodules using ultrasound cine-loop video sequences. Cancers (Basel) 2025;17(16):2616; doi: 10.3390/cancers17162616.

12. Rosenbaum L, Freesmeyer M, Schmidt TN, et al. Ultrasound risk stratification of autonomously functioning thyroid nodules: cine loop video sequences versus static image captures. Diagnostics (Basel) 2025;15(19):2525; doi: 10.3390/diagnostics15192525.

13. Cibas ES, Ali SZ. The 2017 Bethesda System for Reporting Thyroid Cytopathology. Thyroid 2017;27(11):1341–1346; doi: 10.1089/thy.2017.0500.

14. Buda M, Wildman-Tobriner B, Hoang JK, et al. Management of thyroid nodules seen on US images: deep learning may match performance of radiologists. Radiology 2019;292(3):695–701; doi: 10.1148/radiol.2019181343.

15. Cohen J. A coefficient of agreement for nominal scales. Educ Psychol Meas 1960;20(1):37–46; doi: 10.1177/001316446002000104.

16. Fleiss JL. Measuring nominal scale agreement among many raters. Psychol Bull 1971;76(5):378–382; doi: 10.1037/h0031619.

17. Obuchowski NA, Rockette HE. Hypothesis testing of diagnostic accuracy for multiple readers and multiple tests: an ANOVA approach with dependent observations. Commun Stat Simul Comput 1995;24(2):285–308; doi: 10.1080/03610919508813243.

18. Hillis SL. A comparison of denominator degrees of freedom methods for multiple observer ROC analysis. Stat Med 2007;26(3):596–619; doi: 10.1002/sim.2532.

19. Rutter CM. Bootstrap estimation of diagnostic accuracy with patient-clustered data. Acad Radiol 2000;7(6):413–419; doi: 10.1016/S1076-6332(00)80381-5.

20. Hoenig JM, Heisey DM. The abuse of power: the pervasive fallacy of power calculations for data analysis. Am Stat 2001;55(1):19–24; doi: 10.1198/000313001300339897.

21. Peng S, Liu Y, Lv W, et al. Deep learning-based artificial intelligence model to assist thyroid nodule diagnosis and management: a multicentre diagnostic study. Lancet Digit Health 2021;3(4):e250–e259; doi: 10.1016/S2589-7500(21)00041-8.

22. Chi J, Walia E, Babyn P, et al. Thyroid nodule classification in ultrasound images by fine-tuning deep convolutional neural network. J Digit Imaging 2017;30(4):477–486; doi: 10.1007/s10278-017-9997-y.

23. U.S. Food and Drug Administration. 510(k) Premarket Notification K203502: MEDO-Thyroid (MEDO DX Pte. Ltd.); decision date April 23, 2021. Available from: https://www.accessdata.fda.gov/cdrh_docs/pdf20/K203502.pdf [Last accessed: October 14, 2026].

24. U.S. Food and Drug Administration. 510(k) Premarket Notification K240036: PIUR tUS Infinity (PIUR Imaging GmbH); decision date September 20, 2024. Available from: https://www.accessdata.fda.gov/cdrh_docs/pdf24/K240036.pdf [Last accessed: October 14, 2026].

25. Liu D, Yang K, Zhang C, et al. Fully-automatic detection and diagnosis system for thyroid nodules based on ultrasound video sequences by artificial intelligence. J Multidiscip Healthc 2024;17:1641–1651; doi: 10.2147/JMDH.S439629.

26. Yamashita R, Kapoor T, Alam MN, et al. Toward reduction in false-positive thyroid nodule biopsies with a deep learning–based risk stratification system using US cine-clip images. Radiol Artif Intell 2022;4(3):e210174; doi: 10.1148/ryai.210174.

27. Ni C, Feng B, Yao J, et al. Value of deep learning models based on ultrasonic dynamic videos for distinguishing thyroid nodules. Front Oncol 2022;12:1066508; doi: 10.3389/fonc.2022.1066508.

28. Leeflang MM, Rutjes AW, Reitsma JB, et al. Variation of a test's sensitivity and specificity with disease prevalence. CMAJ 2013;185(11):E537–E544; doi: 10.1503/cmaj.121286.

29. Middleton WD, Teefey SA, Reading CC, et al. Comparison of performance characteristics of American College of Radiology TI-RADS, Korean Society of Thyroid Radiology TIRADS, and American Thyroid Association guidelines. AJR Am J Roentgenol 2018;210(5):1148–1154; doi: 10.2214/AJR.17.18822.

30. Hoang JK, Middleton WD, Farjat AE, et al. Reduction in thyroid nodule biopsies and improved accuracy with American College of Radiology Thyroid Imaging Reporting and Data System. Radiology 2018;287(1):185–193; doi: 10.1148/radiol.2018172572.

31. Kim PH, Suh CH, Baek JH, et al. Diagnostic performance of four ultrasound risk stratification systems: a systematic review and meta-analysis. Thyroid 2020;30(8):1159–1168; doi: 10.1089/thy.2019.0812.

Table 1. Nodule features as assessed by the 4 readers across 3 rounds in total. Values are numbers of reader assessments (4 readers × 100 nodules = 400 assessments per round); echogenic foci categories are not mutually exclusive.

| | Round 1 | Round 2 | Round 3 |
|---|---|---|---|
| **Composition** | | | |
| Cystic | 6 | 8 | 6 |
| Spongiform | 0 | 3 | 2 |
| Mixed cystic/solid | 53 | 44 | 56 |
| Solid | 332 | 334 | 324 |
| Cannot be determined | 9 | 11 | 12 |
| **Echogenicity** | | | |
| Anechoic | 3 | 1 | 2 |
| Hyper/isoechoic | 162 | 163 | 176 |
| Hypoechoic | 171 | 165 | 168 |
| Very hypoechoic | 55 | 56 | 40 |
| Cannot be determined | 9 | 15 | 14 |
| **Shape** | | | |
| Wider-than-tall | 314 | 313 | 310 |
| Taller-than-wide | 86 | 87 | 90 |
| **Margin** | | | |
| Smooth | 237 | 207 | 251 |
| Ill-defined | 92 | 117 | 70 |
| Lobulated/irregular | 55 | 63 | 65 |
| Extrathyroidal extension | 8 | 4 | 10 |
| Cannot be determined | 8 | 9 | 4 |
| **Echogenic Foci** | | | |
| None or large comet tail | 252 | 228 | 233 |
| Macrocalcifications | 68 | 65 | 94 |
| Peripheral (rim) calc. | 28 | 32 | 42 |
| Punctate echogenic foci | 96 | 109 | 101 |

Table 2. Sensitivity and Specificity of the FNA recommendation for all four readers in the three rounds (95% confidence intervals in [])

| | | Round 1 - stills only | Round 2 - stills only | Round 3 - cine included |
|---|---|---|---|---|
| Reader 1 | sensitivity | 0.60 [0.46, 0.73] | 0.66 [0.53, 0.79] | 0.72 [0.59, 0.84] |
| | specificity | 0.12 [0.04, 0.22] | 0.20 [0.10, 0.32] | 0.22 [0.11, 0.34] |
| Reader 2 | sensitivity | 0.66 [0.53, 0.79] | 0.68 [0.55, 0.80] | 0.76 [0.64, 0.87] |
| | specificity | 0.26 [0.14, 0.38] | 0.20 [0.10, 0.32] | 0.20 [0.10, 0.32] |
| Reader 3 | sensitivity | 0.68 [0.55, 0.81] | 0.66 [0.53, 0.79] | 0.64 [0.50, 0.77] |
| | specificity | 0.18 [0.08, 0.29] | 0.24 [0.13, 0.36] | 0.22 [0.11, 0.34] |
| Reader 4 | sensitivity | 0.60 [0.46, 0.73] | 0.62 [0.48, 0.75] | 0.56 [0.42, 0.70] |
| | specificity | 0.20 [0.09, 0.32] | 0.16 [0.07, 0.27] | 0.24 [0.13, 0.36] |
| Average | sensitivity | 0.64 [0.52, 0.75] | 0.66 [0.54, 0.76] | 0.67 [0.56, 0.77] |
| | specificity | 0.19 [0.10, 0.28] | 0.20 [0.11, 0.30] | 0.22 [0.13, 0.32] |

Table 3. Number of cases with changed management decisions (FNA vs no FNA) between the rounds for all readers (95% confidence intervals in [])

| | Round 1 vs Round 2 (stills only) | Round 1 vs Round 3 (stills vs stills+cine) | Round 2 vs Round 3 (stills vs stills+cine) |
|---|---|---|---|
| Reader 1 | 13 [7, 20] | 15 [8, 22] | 10 [5, 16] |
| Reader 2 | 10 [5, 16] | 14 [8, 21] | 8 [3, 14] |
| Reader 3 | 12 [6, 19] | 12 [6, 19] | 14 [8, 21] |
| Reader 4 | 9 [4, 15] | 14 [8, 21] | 13 [7, 20] |
| Average | 11.0 [7.0, 15.5] | 13.8 [9.8, 18.0] | 11.3 [8.0, 15.0] |

Table 4. Average absolute TI-RADS points difference between each round for all readers (95% confidence intervals in [])

| | Round 1 vs Round 2 | Round 1 vs Round 3 | Round 2 vs Round 3 |
|---|---|---|---|
| Reader 1 | 1.00 [0.74, 1.28] | 1.02 [0.77, 1.29] | 1.10 [0.81, 1.41] |
| Reader 2 | 1.21 [0.95, 1.49] | 1.93 [1.58, 2.30] | 1.76 [1.43, 2.11] |
| Reader 3 | 0.98 [0.73, 1.25] | 1.28 [1.00, 1.58] | 1.32 [1.07, 1.58] |
| Reader 4 | 1.07 [0.81, 1.35] | 1.17 [0.89, 1.47] | 1.50 [1.16, 1.87] |
| Average | 1.07 [0.92, 1.22] | 1.35 [1.20, 1.51] | 1.42 [1.26, 1.59] |

Table 5. Agreement of each TI-RADS feature assignment between each round for all readers (proportion of nodules assigned the same category by the same reader in both rounds)

| Comparison | Reader | Composition | Echogenicity | Shape | Margin | Artifacts | Macrocalcifications | Peripheral Calcifications | Punctate Echogenic Foci | Average Agreement (Reader) | Average Agreement (Round) |
|---|---|---|---|---|---|---|---|---|---|---|---|
| R1-R2 (static vs static) | Reader 1 | 0.95 | 0.75 | 0.92 | 0.75 | 0.91 | 0.95 | 0.98 | 0.91 | 0.89 | |
| | Reader 2 | 0.87 | 0.82 | 0.90 | 0.66 | 0.76 | 0.91 | 0.98 | 0.88 | 0.85 | |
| | Reader 3 | 0.93 | 0.79 | 0.94 | 0.67 | 0.91 | 0.93 | 0.93 | 0.90 | 0.88 | |
| | Reader 4 | 0.94 | 0.79 | 0.93 | 0.79 | 0.80 | 0.88 | 0.95 | 0.84 | 0.87 | 0.87 |
| R1-R3 (static vs cine) | Reader 1 | 0.91 | 0.75 | 0.94 | 0.68 | 0.84 | 0.94 | 0.97 | 0.90 | 0.87 | |
| | Reader 2 | 0.93 | 0.74 | 0.86 | 0.57 | 0.76 | 0.80 | 0.89 | 0.76 | 0.79 | |
| | Reader 3 | 0.92 | 0.68 | 0.89 | 0.61 | 0.81 | 0.86 | 0.94 | 0.82 | 0.82 | |
| | Reader 4 | 0.89 | 0.83 | 0.91 | 0.81 | 0.80 | 0.82 | 0.96 | 0.85 | 0.86 | 0.83 |
| R2-R3 (static vs cine) | Reader 1 | 0.92 | 0.73 | 0.92 | 0.64 | 0.87 | 0.93 | 0.99 | 0.89 | 0.86 | |
| | Reader 2 | 0.86 | 0.76 | 0.86 | 0.64 | 0.72 | 0.81 | 0.91 | 0.74 | 0.79 | |
| | Reader 3 | 0.89 | 0.81 | 0.87 | 0.64 | 0.82 | 0.87 | 0.93 | 0.84 | 0.83 | |
| | Reader 4 | 0.88 | 0.79 | 0.92 | 0.76 | 0.78 | 0.80 | 0.95 | 0.75 | 0.83 | 0.83 |

Table 6. Agreement statistics for feature assignments. Intra: intrareader agreement, Cohen's κ between two rounds for the same reader (mean of four readers, matched by nodule); Inter: interreader agreement, Fleiss' κ among the four readers within a round.

| | **Intra R1-R2** | **Intra R1-R3** | **Intra R2-R3** | **Inter R1** | **Inter R2** | **Inter R3** |
|---|---|---|---|---|---|---|
| Composition | 0.74 | 0.71 | 0.64 | 0.67 | 0.65 | 0.69 |
| Echogenicity | 0.65 | 0.59 | 0.63 | 0.52 | 0.39 | 0.49 |
| Shape | 0.73 | 0.66 | 0.65 | 0.56 | 0.50 | 0.47 |
| Margin | 0.51 | 0.39 | 0.43 | 0.28 | 0.37 | 0.30 |
| Artifacts | 0.68 | 0.59 | 0.58 | 0.56 | 0.54 | 0.47 |
| Macrocalcifications | 0.71 | 0.57 | 0.55 | 0.67 | 0.63 | 0.52 |
| Peripheral Calcifications | 0.70 | 0.65 | 0.68 | 0.67 | 0.66 | 0.59 |
| Punctate Echogenic Foci | 0.67 | 0.54 | 0.49 | 0.43 | 0.42 | 0.48 |

Table 7. Discrimination and size-independent performance of TI-RADS in each round (mean of four readers). TR4 or higher and TR5 denote TI-RADS levels used as the positive test without size criteria; the FNA recommendation rows (with size criteria, as in Table 2) are shown for comparison. AUC comparisons (Obuchowski–Rockette, readers and nodules random): round 1 vs round 3, difference 0.022 [95% CI: −0.025, 0.070], p=0.30; round 2 vs round 3, difference 0.017 [95% CI: −0.021, 0.055], p=0.38; round 1 vs round 2, difference 0.005 [95% CI: −0.052, 0.063], p=0.80.

| | **Round 1 - stills only** | **Round 2 - stills only** | **Round 3 - cine included** |
|---|---|---|---|
| AUC of TI-RADS point totals | 0.623 | 0.628 | 0.645 |
| Sensitivity, TR4 or higher | 0.82 | 0.83 | 0.83 |
| Specificity, TR4 or higher | 0.29 | 0.31 | 0.35 |
| Sensitivity, TR5 | 0.43 | 0.46 | 0.49 |
| Specificity, TR5 | 0.81 | 0.73 | 0.73 |
| Sensitivity, FNA recommendation | 0.64 | 0.66 | 0.67 |
| Specificity, FNA recommendation | 0.19 | 0.20 | 0.22 |